# Self-organized Kagomé-lattice in a metal-organic monolayer

*Nesrine Shaiek* [1,2], *Hassan Denawi* [1,3], *Mathieu Koudia* [1], *Roland Hayn* [1], *Steffen Schäfer* [1], *Isabelle Berbezier* [1], *Chokri Lamine* [2], *Olivier Siri* [4], *Abdelwaheb Akremi* [2] *and Mathieu Abel* [1]

[1] Aix Marseille Université, CNRS, IM2NP, UMR 7334, Campus de St Jérôme, 13397, Marseille, France

[2] Université de Carthage, Faculté des Sciences de Bizerte, Laboratoire de Physique des Matériaux: Structure et Propriétés, LR01 ES15, Unité de Service Commun Spectromètre de Surfaces, 7021 Bizerte, Tunisie

[3] Centre d'élaboration de matériaux et d'études structurales (CEMES), CNRS, Université de Toulouse, Toulouse, France

[4] Aix Marseille Université, CNRS, CINAM, UMR 7325, Campus de Luminy, 13288, Marseille, France

## Abstract:

We report on the successful on-surface synthesis of metal-organic covalent coordination networks with a dense Kagomé lattice of metallic centers. In the case of Mn centers ab-initio calculations show that the adsorbed monolayer on Ag(111) has all the characteristic features of a strictly two-dimensional (2D) ferromagnetic Kagomé metal. Tetrahydroxyquinone (THQ) and metal atoms (M=Cu or Mn) are co-deposited on the Ag(111) substrate to build well-ordered 2D lattices $M_3C_6O_6$. The surface is studied by scanning tunneling microscopy (STM), low energy electron diffraction (LEED) and X-ray photoelectron spectroscopy (XPS) to optimize the growth conditions like fluxes and temperatures. The details of the atomic, electronic and magnetic structures are clarified by density functional theory (DFT) calculations. XPS and DFT reveal a $Cu^+$ charge state and no local magnetic moments for the Cu-organic network. For the Mn-organic network, we find the charge state $Mn^{2+}$ and a local spin $S = 5/2$. Charge transfer stabilizes the $Cu^+$ and $Mn^{2+}$ charge states. We find two different modifications of the $M_3C_6O_6$ lattice since each Mn has 4 equivalent oxygen neighbors whereas Cu has two oxygen neighbors closer than the other two. DFT calculations which neglect the small spin-orbit coupling show a Dirac point, i.e. a band crossing with linear electron dispersion at the K-point (2/3) $\vec{g}_a$ + (1/3) $\vec{g}_b$ of the Brillouin zone. This Dirac point is at the Fermi level if there is no charge transfer but drops by 100 meV if electron doping of $Cu_3C_6O_6$ on Ag(111) surface is acknowledged. We predict the magnetic couplings of an isolated $Mn_3C_6O_6$ monolayer to be short range and antiferromagnetic leading to high frustration at the Kagomé lattice and a tendency towards a spin-liquid ground state. In the case of hole transfer from the substrates ferromagnetic ordering is introduced, making $Mn_3C_6O_6$ an interesting candidate for the quantum anomalous Hall effect.

### Introduction

The discovery of graphene by exfoliation of graphite in 2004 [1] initiated the experimental study of truly two-dimensional (2D) materials. Since then, 2D materials have become an extremely fruitful field of research, with hundreds of newly isolated or synthesized artificial monolayers each year, and a multitude of techniques like e.g. molecular beam epitaxy, chemical deposition on surfaces, synthesis in wet environment and many others. Initially mainly studied for their fundamentally new and exotic properties, two-dimensional materials (2D) have now attracted widespread interest due to their high potential for novel nanotechnological applications. [2,3] In this context, 2D materials composed of metal atoms and organic linkers have emerged as promising candidates to combine the advantages of both realms, also called 2D metal organic frameworks (MOFs). For potential applications, we may cite, for example, their abilities in molecular recognition and functionalities for heterogeneous asymmetric catalysis, while their peculiar topological properties are interesting for new spintronic devices. [4-14] Most of the 2D metal organic frameworks (MOFs) synthesized so far are almost perfect insulators but there is a major exception, namely the subclass of conductive MOFs or c-MOFs which are either metallic or at least semi-conducting. The development of these c-MOFs is an important challenge for future decades due to their novel electronic, optical, mechanical, and catalytic properties. [15-19]

Graphene is of fundamental interest since its linearly dispersing highly mobile electrons may be viewed as a 2D version of massless Dirac fermions. But in graphene local magnetic moments are missing which excludes all interesting applications depending on magnetic effects or spin-dependent transport. Here, we report on the successful on-surface synthesis of a dense 2D metal-organic covalent coordination network with a Kagomé lattice of metal ions, thus opening a window to the unconventional electronic and magnetic properties of Kagomé lattices. So, the Kagomé lattice antiferromagnet shows an unconventional spin chirality on a 2D frustrated lattice. [20-22] On the electronic side, the Kagomé lattice combines Dirac fermions with flat bands. The combination with ferromagnetic order then allows for the quantum anomalous Hall (QAH), i.e. a quantum Hall effect without an external magnetic field. This was recently observed in the quasi-2D Kagomé metal $Fe_3Sn_2$, [23] a metallic compound with ferromagnetic order and a band crossing at the highly symmetric K point slightly below the Fermi level. In a ferromagnetic material with only one spin species the band crossing is also called Weyl point (what was the Dirac point in graphene) with concomitant massless quasiparticles. The spin orbit coupling - which is crucial for the QAH effect - opens a gap at the crossing point and leads to the observed massive quasiparticles in $Fe_3Sn_2$. [23] The QAH effect was also predicted for 2D metal-organic systems. [24] In the present paper, we discuss a possible route to obtain that goal in synthesizing a metal-organic monolayer whose electronic structure possesses all the required features for the QAH to occur.

In the search for novel states of matter, the occurrence of a Kitaev spin liquid was predicted in a metal-organic magnetic system with a honeycomb lattice. [25] (Such metal-organic honeycomb lattices were indeed synthesized, but rather to look for applications as c-MOF. [26,27]) The spin-liquid state can also be obtained on the Kagomé lattice with antiferromagnetic (AFM) nearest-neighbor couplings like e.g. in the naturally occurring quasi-2D Herbertsmithite.[28] In this article, we report on the successful synthesis of a $S = 5/2$ Kagomé lattice arising as a self-organized metal-organic monolayer on the Ag(111) surface. Investigating the magnetic couplings by ab-initio calculations we find the interesting phenomena that these couplings can be tuned, with antiferromagnetic (AFM) couplings arising for an isolated layer, and ferromagnetic couplings for the layer on the substrate. To synthesize such a Kagomé lattice, tetrahydroxyquinone (THQ) is particularly attractive because of its ability to form stable complexes with various metal ions in a variety of coordination modes using its four hydroxyl groups.[29,30] Here, we report the surface reaction between transition metal atoms (copper Cu and manganese Mn) and THQ on silver Ag (111) under ultra-high vacuum conditions and at an appropriate temperature - which contrasts with the previously reported synthesis of the c-MOF $Cu_3(C_6O_6)_2$ in aqueous solution. [31] The deposition of $C_6O_6H_6$ on Cu(111) with an appropriate thermal treatment leads to a dense 2D $Cu_3C_6O_6$ lattice [32] whose properties differ significantly from the traditional porous MOFs. In our lattice, the Cu ions form a Kagomé network, but similar lattices were synthesized with Fe. [33,34] The adsorption of THQ molecules alone, [35] without metallic linker, leads to an ordered arrangement of organic molecules with no chemical reaction, i.e. no real chemical binding. The organic molecules stack together only via hydrogen bonds.

Here, we report on the results of co-deposition of two transition metals (Cu and Mn) with $C_6O_6H_4$ on the Ag(111) surface (**Figure 1**). Using metallic linkers, we obtain two different, chemically stable 2D metal-organic covalent coordination networks. The deposition temperature is chosen such that the metals adsorb easily but not the organic molecules. The latter can only adsorb afterwards, by forming chemical bonds with the adsorbed metals. The 4 bonds around each Mn are identical in $Mn_3C_6O_6$, whereas 2 oxygens are closer than the other 2 in $Cu_3C_6O_6$ due to a rotation of the $C_6O_6$ rings, resulting in a considerable larger lattice constant for the Cu-network. The different lattice constants become obvious from the Moiré pattern which is observed for the Cu network but absent with manganese since the $Mn_3C_6O_6$ lattice matches perfectly with that of Ag (111). During the chemical binding process the organic molecules lose all their hydrogens and build a c-MOF, which could in principle be removed from the metallic substrate.

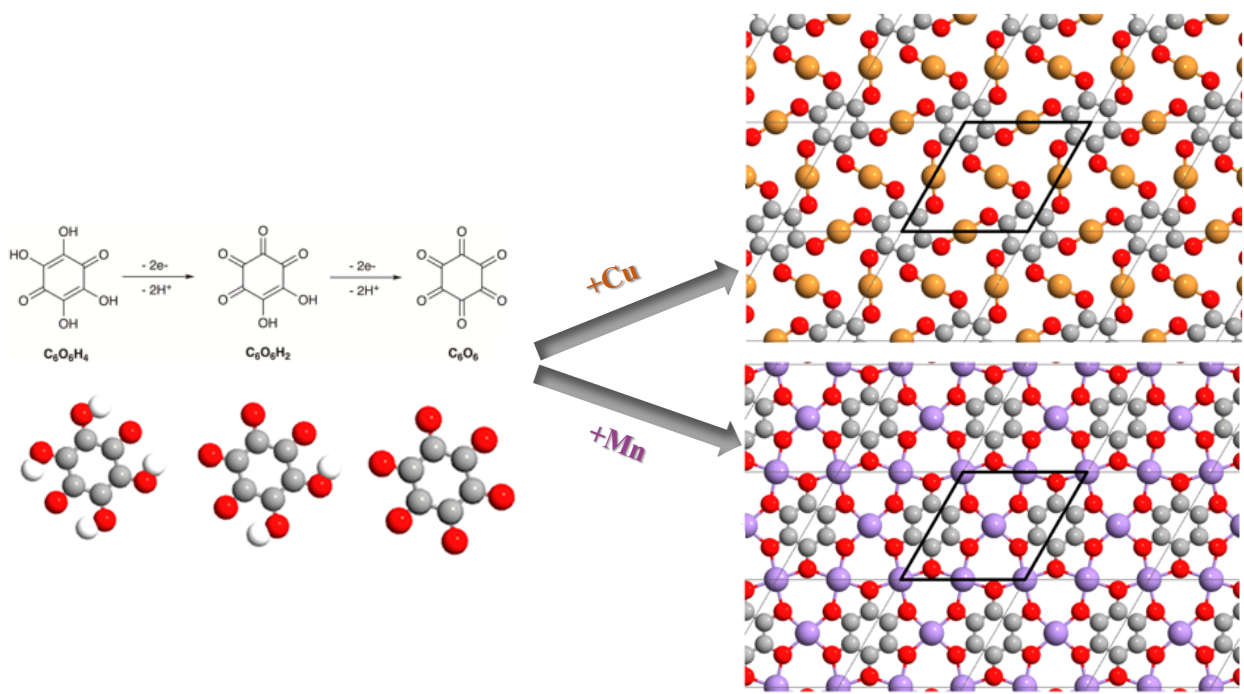

**Figure 1.** On-surface synthesis of THQ (tetrahydroxyquinone) together with Mn or Cu to dense 2D c-MOFs $Mn_3C_6O_6$ and $Cu_3C_6O_6$ (O-red, C-grey, Cu-orange, and Mn-purple).

**Experimental section**

Experiments are performed in an ultra-high vacuum (UHV) system from Omicron instruments. The metals Cu and Mn are introduced from an EFM evaporator while the ligand precursor tetrahydroxyquinone THQ (Aldrich) is deposited via an evaporator crucible held at 410 K and is let to react on the Ag(111) surface with the metal atoms. The formation of a metal-organic coordination network on the surface is further supported by scanning tunneling microscopy (STM), low energy electron diffraction (LEED) and X-ray photoemission spectroscopy (XPS) measurements. **Figure 2a** displays the STM image recorded after deposition of THQ on Ag(111) at room temperature. The lattice parameter is 7.65 Å in agreement with a previous study. [35] The LEED diagram (**Figure 2b**) indicates the well-known ($\sqrt{7}\times\sqrt{7}$) superstructure of THQ on Ag(111) rotated by ±20° with respect to the Ag(111) close packed directions.

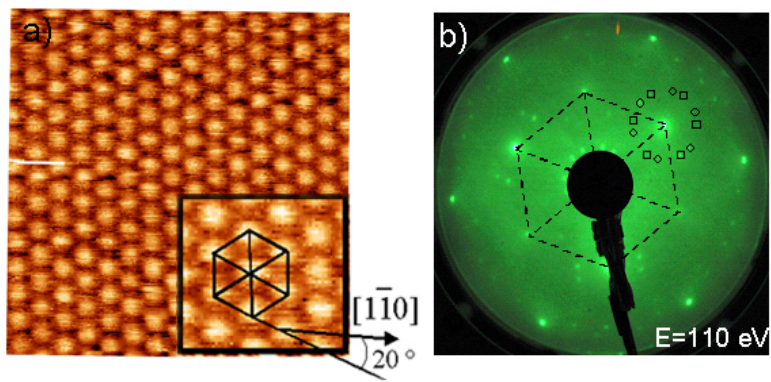

**Figure 2.** THQ monolayer deposited on Ag(111) at room temperature **(a)** (10 nm x 10 nm) STM image **(b)** LEED pattern of the ($\sqrt{7}\times\sqrt{7}$)R20° superstructure (110 eV) . [35]

The co-deposition of THQ molecules and Cu atoms on the Ag (111) surface at 150 °C allows the formation of large and well-ordered domains of the metal-organic coordination network whereas the desorption temperature of the THQ molecule alone is around 120°C. The deposition of THQ molecules on the substrate kept at 150°C is only possible thanks to the presence of the co-deposited Cu atoms. The STM image (**Figure 3a**) shows a well-organized domain of (8.17 ± 0.4) Å with a very slight modulation of the corrugation. This can be rationalized as a Moiré pattern with a lattice parameter of 5.5 nm ± 0.2 nm oriented at 10° from the [1-10] direction. Within these domains the molecules are organized in a centered hexagonal network with 6 copper atoms surrounding each molecule. The mesh parameter of this lattice obtained by LEED is (8.10 ± 0.2) Å

**(Figure 3c)**. It is important to note that this lattice parameter is larger than the ($\sqrt{7} \times \sqrt{7}$) lattice substrate (i.e. 7.65 Å). This is at the origin of the appearance of a large-scale coincidence (5.5 nm Moiré) resulting from a $Cu_3C_6O_6$ lattice parameter of 8.056 Å (see Supplemental Information (SI) S5).

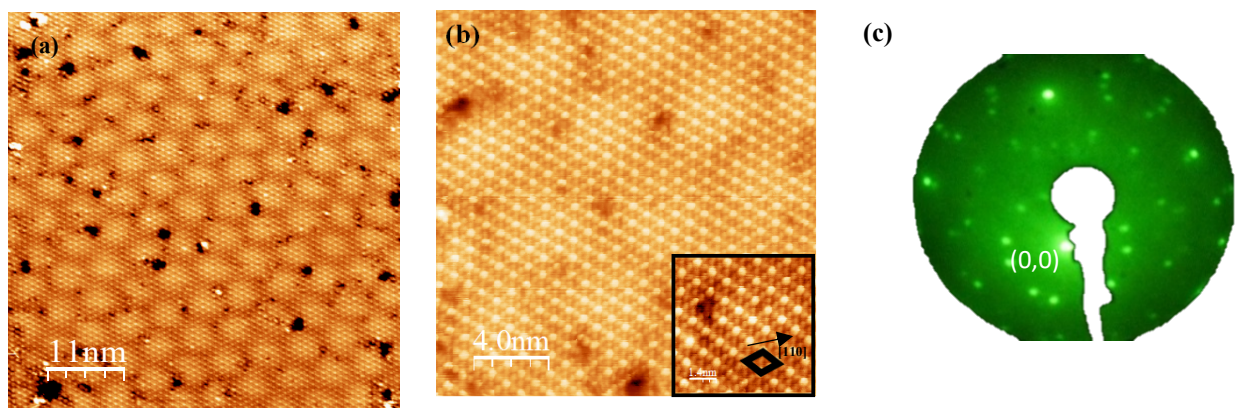

**Figure 3.** Structural domains of THQ molecules and copper atoms co-deposited on Ag(111) at 150°C. (a) and (b) STM Images of 55 nm × 55 nm and 20 nm x 20 with inset (7 nm × 7 nm), respectively. (c) LEED pattern.

The XPS measurements **(Figure 4)** allow to determine the chemical state of the different species. Core level spectra of O 1s, C 1s and Cu 2p3/2 were recorded and compared to as-deposited molecules and copper atoms alone to confirm the reaction between the copper atoms and THQ. This indicates that all the molecules are dehydrogenated as demonstrated by Zhang et al. in the case of $C_6O_6H_6$ on Cu(111) and that the copper atoms are in a Cu(I) (or Cu$^+$) configuration.[32]

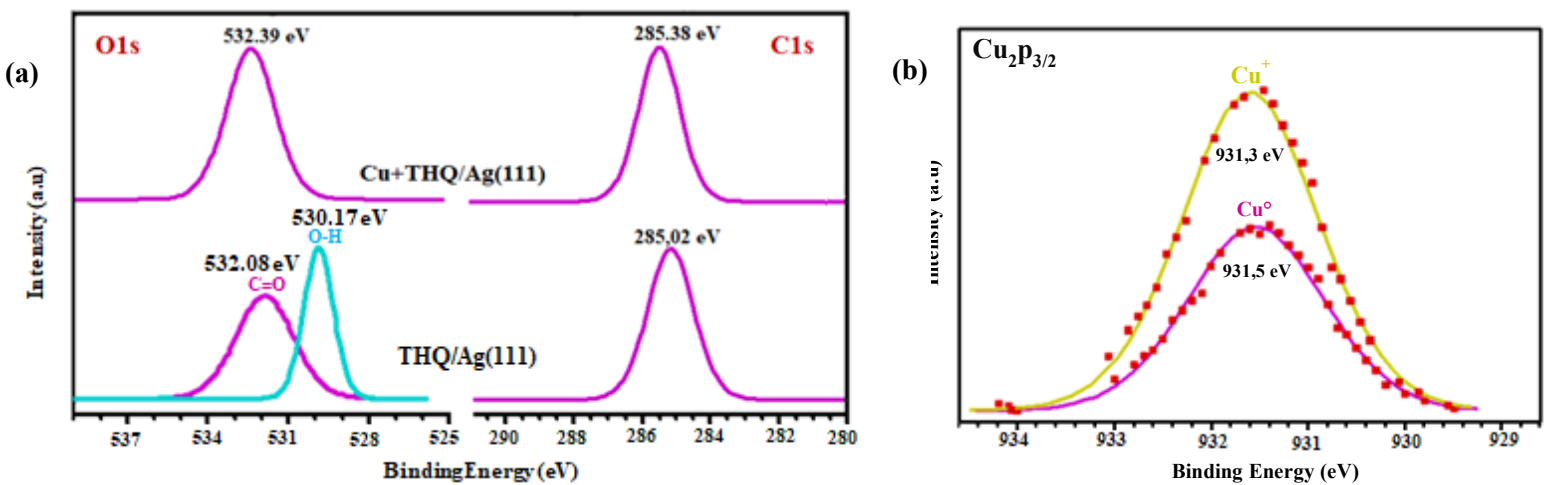

**Figure 4.** Core levels of (a) O 1s and C 1s for THQ and co-deposited Cu and THQ on Ag(111), (b) Cu 2p3/2 for Cu on Ag(111) (red line) and Cu and THQ on Ag(111) (yellow line).

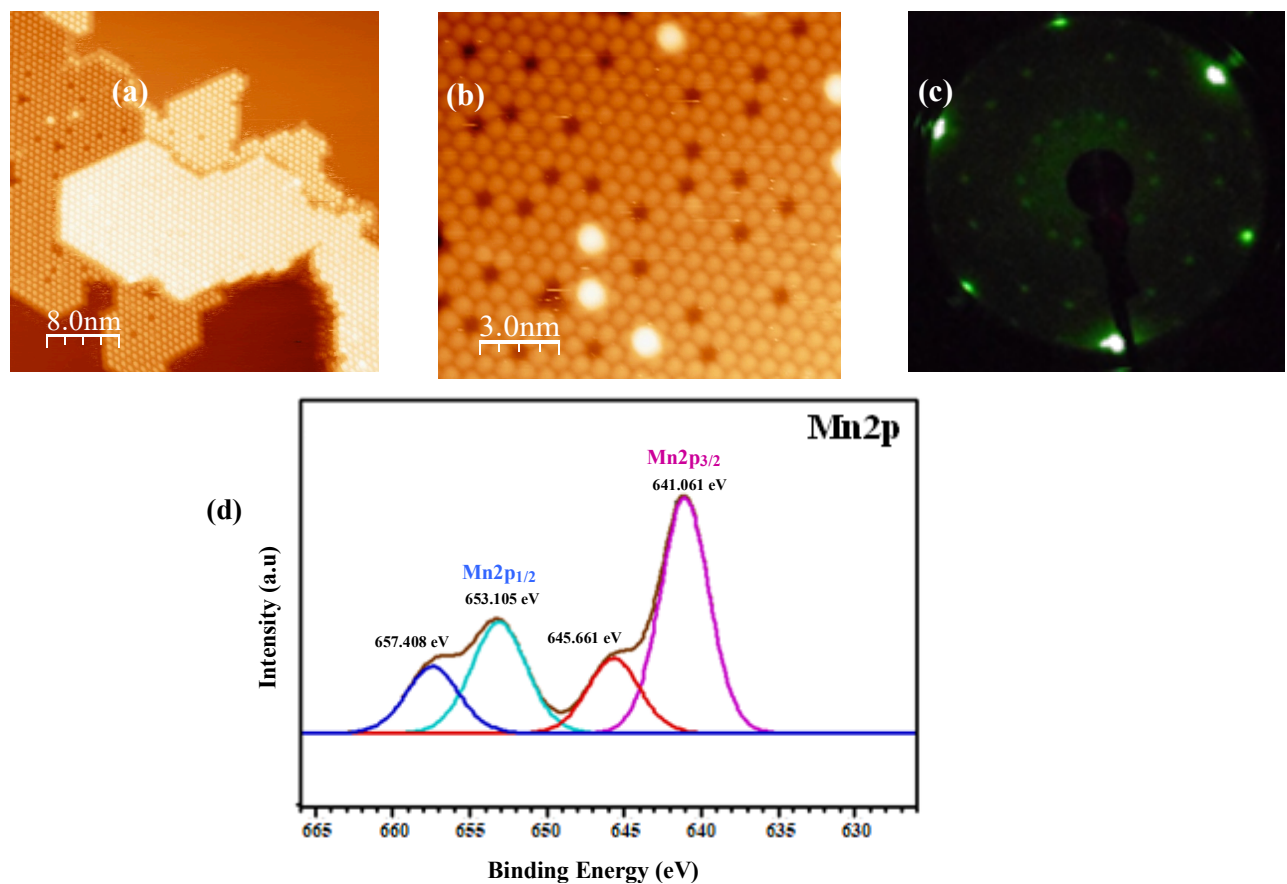

**Figure 5.** Structural domains of THQ molecules and manganese atoms co-deposited on Ag (111) at 150°C. (a) and (b) STM images of (40 nm × 40 nm) and (15 nm × 15 nm), respectively. (c) LEED pattern. (d) Mn 2p XPS spectrum of the THQ-Mn 2D MOF after co-depositing at 200°C.

In the case of THQ and Mn co-deposition, the Moiré pattern is not present and all the molecules are on equivalent adsorption sites thanks to a smaller mesh size than in the previous case with copper. As deduced from the STM and LEED data the lattice parameter is (7.67 $\pm$0.2) Å **(Figure 5)** corresponding to the ($\sqrt{7}\times\sqrt{7}$)R20°. The deconvolution of XPS spectrum of Mn 2p **(Figure 5d)** at 641.06 eV and 653.11 eV indicates a Mn(II) (or Mn$^{2+}$) configuration when compared to Mn deposition alone on Ag (111).[36]

**Theoretical results**

To supplement and interpret the experimental findings, several ab-initio calculations have been performed. We used the Vienna Ab-initio Simulation Package (VASP) and density functional theory in the form of the spin unpolarized generalized gradient approximation (GGA) for Cu$_3$C$_6$O$_6$, and using the spin-polarized version (SGGA) for Mn$_3$C$_6$O$_6$. We also investigated the influence of the Hubbard $U$ correction (SGGA+U functional) on the incompletely filled 3d shell which is found in Mn$_3$C$_6$O$_6$ (see SI for more details on the method). We modeled isolated M$_3$C$_6$O$_6$ layers (M=Cu or Mn) in two different configurations as a function of the lattice constant, as well as adsorbed layers on the Ag(111) surface of the substrate.

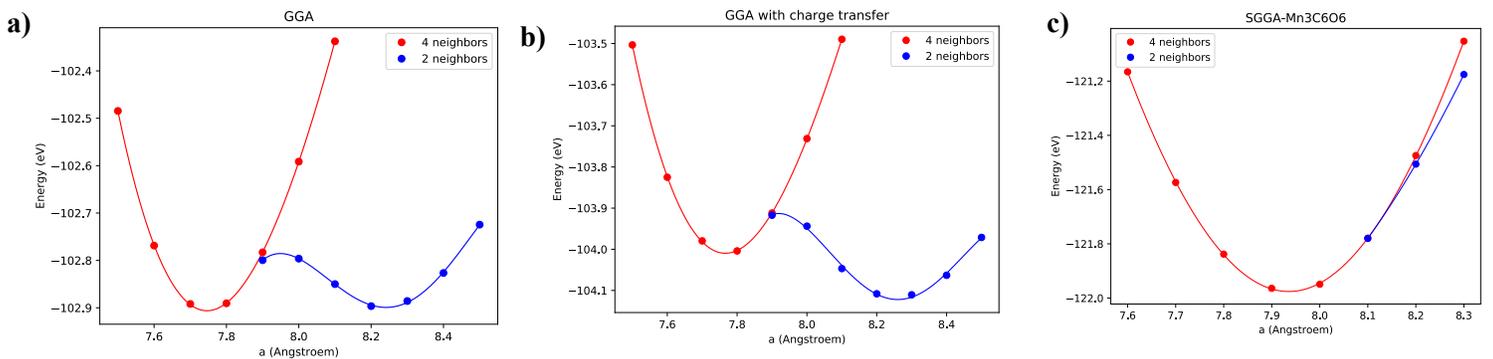

**Figure 6.** Total energies as a function of lattice constant for Cu$_3$C$_6$O$_6$ without (a) and with charge transfer (b), as well as for Mn$_3$C$_6$O$_6$ (panel c) using GGA or SGGA.

Let us first discuss the results for isolated monolayers and start with Cu$_3$C$_6$O$_6$ **(Figure 6a)**. Here, we show results by using the GGA functional without spin splitting. It should be remarked that there exists a ferromagnetic solution which is nearly degenerate with the nonmagnetic one. As will be explained below, this ferromagnetic solution is not relevant for $Cu_3C_6O_6$ since it becomes unstable due to the charge transfer from the Ag substrate. The crystal structure where each Cu would have 4 equivalent O neighbors (which we will denote as structure A) is that one which is realized for $Mn_3C_6O_6$ as shown in **Figure 1**. It leads to the energy minimum at 7.75 Å of the red curve in **Figure 6a**. However, the Bader charge analysis, whose details are given in the Supplemental Information (SI), points to a completely filled d-shell, implying that the GGA solution should display a copper charge closer to $Cu^+$ than to $Cu^{2+}$. In analogy to the difference between the crystal structures of CuO and $Cu_2O$, it can be expected that $Cu^+$ will prefer just two oxygen neighbors over a configuration with four. This preference can be met by rotating the $C_6O_6$ rings while keeping the Kagomé lattice for the metal ions fixed (oxygen ions are so tightly bound to the $C_6$ ring that they will rotate with the carbon ring). In the following, we will denote as structure B the configuration with only two O neighbors as depicted in **Figure 1** for $Cu_3C_6O_6$. Ab-initio calculations allow to distinguish between structures A and B by differently choosing the initial atomic positions. For structure B, the rotation angle of the $C_6O_6$ rings converges towards an optimal angle for each fixed lattice constant. At that lattice constant where the curve B touches curve A both solutions coincide. It is clearly visible in **Figure 6a** that there is a second energy minimum for structure B, with a larger lattice constant of 8.25 Å, which is energetically almost degenerate with structure A. Our result coincides with previous ab initio calculations.[32]

The situation for $Mn_3C_6O_6$ is completely different. First of all, the magnetic solution is several electron volts deeper in energy than the nonmagnetic one. That is visible in **Table 1**, where we present the energy difference between the high spin solution ($S = 5/2$) and the intermediate spin solution ($S = 3/2$). The difference with the non-magnetic one is even larger. We also see, that the inclusion of the Hubbard $U$ correction prefers the high spin solution even more and leads to the appearance of a gap in the electronic spectrum. Furthermore, only structure A (4 oxygen neighbors) is stable (see **Figure 6c**) and there is no second minimum for structure B. For larger lattice constants, structure A simply goes over into structure B and there is no second minimum. Without the Hubbard $U$ correction, we find half-metallic behavior with a magnetic moment of $M = 5\mu_B$. This corresponds to $S = 5/2$, a spin state which is only possible in the case of $Mn^{2+}$ oxidation state. However, in analogy to ab-initio results for similar metal-organic monolayers [37, 38] or for other 3d transition metal compounds in general, we do not believe in the half-metallic behavior of the simple SGGA functional. On the contrary, we think the SGGA+U functional, which gives a gap of 1.22 eV for a $U$ parameter of 5 eV, to be much more realistic. The magnetic state ($S = 5/2$) does not change by including the Hubbard $U$ correction. In the SI we compare also results between $U = 4$ and 5 eV showing that a small change of $U$ is not essential and we give there more details on the electronic structures of $Cu_3C_6O_6$ and $Mn_3C_6O_6$.

**Table 1.** Energy difference $\Delta_{E_{HS}-E_{IS}}$ between high spin ($S$=5/2) and intermediate spin ($S$=3/2) solutions for isolated monolayers of $Mn_3C_6O_6$ within the SGGA ($U$=0 eV) and SGGA+U functionals. Also given are the distance between Mn and O ($d_{Mn-O}$), the lattice constant ($a$), the exchange energies ($E_{ex-1/2}=E_{AF(1/2)}-E_{FM}$ per Mn atom) and coupling constants ($J_1$ and $J_2$), as well as the total energy gap ($E_g$).

|  | $U$ = 0 eV | $U$ = 5 eV |
| --- | --- | --- |
| $\Delta_{E_{HS}-E_{IS}}$ (eV) | -1.96 | -4.38 |
| $d_{Mn-O}$ (Å) | 2.10 | 2.10 |
| $a$ (Å) | 7.941 | 7.975 |
| $E_{ex-1}$ (meV) | -26.07 | -6.08 |
| $E_{ex-2}$ (meV) | -46.35 | -10.87 |
| $J_1$ (meV) | 2.43 | 0.57 |
| $J_2$ (meV) | 0.35 | 0.08 |
| $E_g$ (eV) | 0 | 1.22 |

However, simple monolayer calculations are not sufficient to understand the electronic structure of $M_3C_6O_6$ on the Ag-substrate. For that reason, we introduced the Ag substrate in our ab-initio study by 6 Ag layers. After relaxing the atomic positions, we find the structures presented in **Figure 7**. There we show the results for the Mn system where we used the experimental lattice constant of Ag to which the $Mn_3C_6O_6$ monolayer adapts perfectly (in the ($\sqrt{7}$x$\sqrt{7}$) superstructure) leading to a lattice constant of 7.65 Å. Also, we used the SGGA+U functional with $U = 5\ eV$ for $Mn_3C_6O_6$. To take into account the larger lattice constant of the $Cu_3C_6O_6$ monolayer being visible in the Moiré pattern we expanded the underlying Ag substrate such that we reach a lattice constant of 8.10 Å (see SI). The ab-initio simulation of the Moiré pattern would be too difficult for an ab-initio simulation and is practically impossible. Also, we simulate a nonmagnetic state for the Cu system and use the GGA functional.

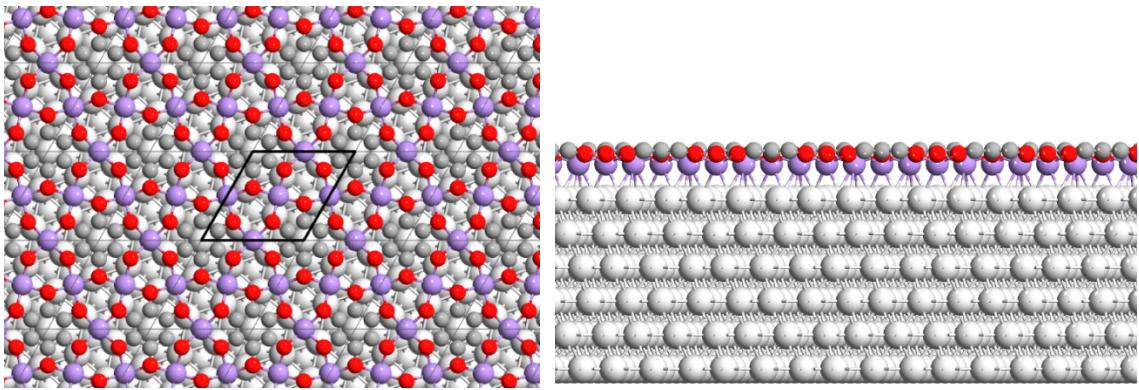

**Figure 7.** Top and side view of $Mn_3C_6O_6$ with substrate as obtained from a SGGA+U calculation with $U$=5 eV.

In the case of the Mn-THQ network **(see Figure 7)** as well for Cu-THQ (see SI) the adsorbed monolayers are not fully planar. The metal ions are closer to the Ag substrate than the $C_6O_6$ rings. More important than these structural details are the charge transfers from the substrate to the $M_3C_6O_6$ (M = Cu or Mn) monolayers. The substrate dopes 0.43 electron to the $Cu_3C_6O_6$ network but removes 0.59 electron from the Mn-organic monolayer. The charge transfer is visible in the Bader charge analysis presented in the SI.

The charge transfer is also the key element to resolve two problems of the monolayer calculation for $Cu_3C_6O_6$ without charge transfer **(Figure 6a)**, namely the near degeneracy between the two minima for structures A and B and between magnetic and nonmagnetic solution. If structures A and B would indeed be energetically degenerate, we would expect phase separation and the presence of phase A also for $Cu_3C_6O_6$. However, as it is presented in **Figure 6b**, taking into account a realistic charge transfer of 0.43 electron doping for the $Cu_3C_6O_6$ monolayer leads to a considerable lowering of the energy for structure B with respect to structure A explaining the experimental findings. Also, the charge transfer stabilizes the nonmagnetic solution with respect to the magnetic one (not shown) and moves the Cu charge closer to $Cu^+$.

Finally, we would like to discuss the band structures close to the Fermi-level which are especially promising for a Kagomé lattice. Since it allows for a band crossing and a linear electron dispersion at the K point in close analogy to the Dirac point in graphene which is at the origin of all the exceptional electronic properties of graphene or other Dirac systems. As we can see in **Figures 8a** and **b**, we predict a similar Dirac point to exist in $Cu_3C_6O_6$ at the K point. It is located exactly at the Fermi-level for the undoped situation but 100 meV below in the electron doped case realized on the Ag(111) substrate.

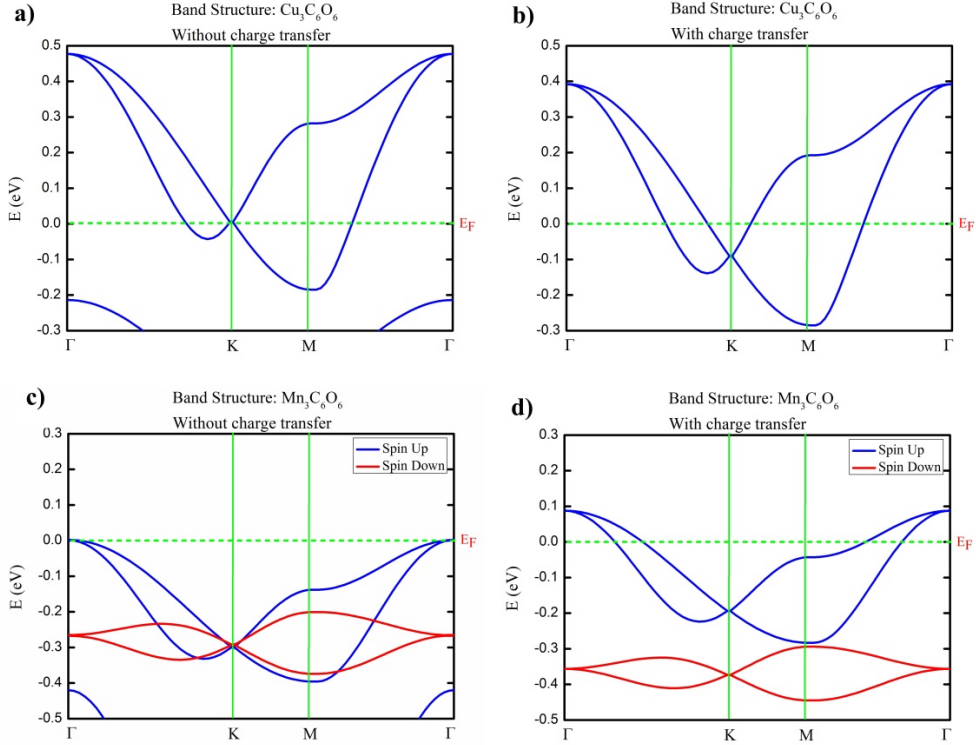

**Figure 8.** Band-structures of $Cu_3C_6O_6$ (non-magnetic) without (a) and with (b) charge transfer of 0.43 electrons, as well as of $Mn_3C_6O_6$ (majority spin-up is black and minority spin-down is red) without (c) and with charge transfer of 0.59 holes from the substrate. The non-magnetic calculations were performed with the GGA functional, and the magnetic ones with SGGA+U functional and $U$=5eV.

In contrast to the metallic behavior of the $Cu_3C_6O_6$ mono-layer (without or with charge transfer) we predict a magnetic, insulating state for $Mn_3C_6O_6$ if there is no charge transfer. However, the hole doping which we discovered in our ab-initio calculation with the Ag substrate leads to a metallic state with the Fermi-level lying in the valence band with majority spin and being of predominantly oxygen character (see **Figure 8** and SI). For this calculation, we did choose a ferromagnetic ordering of the magnetic moment which is the ground-state for the adsorbed monolayer as will be discussed more in detail below. Now, the band crossing at the K point below the Fermi-level has no spin degeneracy and describes Weyl quasiparticles. It should be noted, however, that the inclusion of spin-orbit coupling would open a small gap at the K point.

It is highly interesting that the magnetic moments at the Mn sites build a Kagomé lattice. Let us determine the exchange couplings $J_1$ and $J_2$ to nearest and second nearest neighbors on the Kagomé lattice in the Heisenberg Hamiltonian

$$H = \sum_{\langle i,j \rangle} J_{ij} \vec{S}_i \vec{S}_j = \frac{1}{2} \sum_{i,j} J_{ij} \vec{S}_i \vec{S}_j = \frac{1}{2} \sum_{r,g} J_g \vec{S}_r \vec{S}_{r+g}$$

where $\langle i,j \rangle$ means that we sum over each bond only once. Similar to the linear metal-organic polymer chains [37,38] we take the energy difference between ferromagnetic $E_{FM}$ and antiferromagnetic $E_{AF}$ spin arrangements to calculate $E_{ex} = E_{AF} - E_{FM}$ which we define per magnetic ion. For a one-dimensional (1D) Heisenberg chain with only nearest neighbor exchange $J_1$ and taking the approximation of classical spins $\vec{S}_i \vec{S}_j = \pm S^2$ for parallel or antiparallel arrangement we obtain $E_{ex} = -2J_1 S^2$, where we correct a factor of 2 in the formulas used in [37,38]. (As a consequence of this mistake all the exchange couplings given in [37,38] have to be multiplied by a factor of 2). For a ferromagnetic arrangement of the spins $S = 5/2$ on the Kagomé lattice we obtain $E_{FM} = 2(J_1 + J_2)S^2$ whereas the two antiferromagnetic arrangements (see SI) give

$$E_{AF1} = \frac{2}{3}(J_1 - J_2)S^2 \ ,$$

$$E_{AF2} = \frac{-2}{3}(J_1 + J_2)S^2 \ .$$

Taking the two exchange energies $E_{ex,1} = E_{AF1} - E_{FM}$ and $E_{ex,2} = E_{AF2} - E_{FM}$ for $U = 5 \ eV$ **(see Table 1)** we obtain a considerable antiferromagnetic exchange coupling $J_1 = 0.57 \ meV$ to nearest neighbors but a roughly 10 times smaller (also antiferromagnetic) coupling $J_2 = 0.08 \ meV$ to second neighbors. We give in Table 1 also the $J$ values calculated without $U$ and in the SI the $J$ values with a $U$ of 4 eV.

The exchange couplings change if we take into account the substrate explicitly and consider the adsorbed mono-layer. The energy difference between the two antiferromagnetic solutions remains nearly the same, but the ferromagnetic one goes drastically down in energy such that it becomes the ground-state ($E_{ex,1} = 13.25$ meV and $E_{ex,2} = 10.09$ meV). It means that the exchange coupling to nearest neighbors $J_1 = 0.38$ meV remains antiferromagnetic, but the ferromagnetic second neighbor exchange coupling $J_2 = -0.98$ meV dominates now the magnetic order. The main reason for that transition to a ferromagnetic state is the charge transfer which leads already to $E_{ex,1} = 2.78$ meV and $E_{ex,2} = -2.77$ meV ($J_1 = 0.67$ meV and $J_2 = -0.57$ meV) if we calculate just an isolated monolayer with the 0.59 holes charge transfer. A further stabilization of the ferromagnetic order occurs due to exchange paths which include the Ag substrate.

### Perspectives and discussion

We demonstrated a new synthesis route to achieve a dense 2D metal-organic polymer network of $M_3C_6O_6$ on Ag(111) which was worked out for the metals M=Cu or Mn. But we see no principal reason why it should not work for other metals also beyond the 3d transition metals or other substrates. So, we found a rather general way to synthesize Kagomé lattices of magnetic centers with different spin values. Also, our two examples show that the substrate generally dopes the metal-organic monolayers (electron or hole doping) and may change an insulating monolayer into a metallic one. By choosing different substrates, the doping may eventually be tuned.

We would like to mention a principal difference of our newly synthesized structures to metal-organic frameworks (MOFs) which are already known for several years. These MOFs (not necessarily 2D) are most commonly synthesized in aqueous solution and they are rather porous. In our case, after the on-site polymerization reaction, we obtain a rather dense 2D metal-organic network with small metal-metal distances. For instance, if we compare the MOF $Cu_3 (C_6O_6)_2$ [31] synthesized in aqueous solution with our c-MOF network $Cu_3C_6O_6$ due to on-surface polymerization, we observe in both cases a Kagomé lattice of metal ions. However, the nearest neighbor Cu-Cu distance is 7.0 Å in $Cu_3 (C_6O_6)_2$ reducing to 4.1 Å in the newly synthesized c-MOF network $Cu_3C_6O_6$.

The Cu ions in the c-MOF network $Cu_3C_6O_6$ on Ag (111) are in the $Cu^+$ charge state with a completely filled d-shell and no magnetic moment. The charge transfer of 0.43 electrons from the substrate to the monolayer stabilizes the $Cu^+$ charge state which is comparable with the situation on the Cu (111) surface [32] where the charge transfer was found to be 0.66 electrons. That c-MOF network is interesting as a highly conducting metal-organic layer. Our ab-initio calculations indicate metallic behavior for a free-standing $Cu_3C_6O_6$ monolayer with the Fermi level lying in a band of about 0.7 eV width. Interestingly, we find a Dirac point being exactly located at the Fermi level. The adsorption on Ag(111) leads to a down-shift of this band with respect to the Fermi level as it was also observed for $Cu_3C_6O_6$ on Cu(111). [32]

We find local magnetic moments in $Mn_3C_6O_6$ and metallic behavior due to hole-doping by the Ag(111) substrate with the Fermi-level short above the band-crossing point at K. Without doping, the magnetic interactions are short range and mainly between nearest neighbors. However, doping brings the ferromagnetic spin arrangement down in energy with respect to the antiferromagnetic ones and leads to a ferromagnetic order. So, we predict this Mn-monolayer on Ag(111) to be a ferromagnetic Kagomé metal similar to $Fe_3Sn_2$ and we expect the QAH effect. We propose to check this prediction by transport measurements in a magnetic field. Diminishing the magnetic field, the Hall effect should remain even for zero field.

Without charge transfer, i.e. for a free-standing $Mn_3C_6O_6$ monolayer, the antiferromagnetic (AFM) exchange couplings are clearly dominant. For such a system, the magnetic couplings are highly frustrated on the Kagomé lattice and could lead to spin-liquid (SL) behavior. Please note also that the metal-oxygen network of our network is practically identical to that one realized in Herbertsmithite, a model material for SL behavior. [28] There, the Cu ions build a spin 1/2 Kagomé network whereas we find here a spin 5/2 lattice for $Mn_3C_6O_6$. Apparently, the Cu ions in Herbertsmithite are $Cu^{2+}$ in difference to our case due to different charge transfers. So, our synthesis route opens the way to prepare experimental systems representing AFM Kagomé lattices with different values of spin.

**Acknowledgements**

This work was possible with the High-Performance Computing (HPC) resources of Aix-Marseille University financed by the project Equip@Meso (ANR-10-EQPX-29-01).

# Supporting Information

## Self-organized Kagomé-lattice in a metal-organic monolayer

*Nesrine Shaiek* [1,2], *Hassan Denawi* [1,3], *Mathieu Koudia* [1], *Roland Hayn* [1], *Steffen Schaefer* [1], *Isabelle Berbezier* [1], *Chokri Lamine* [2], *Olivier Siri* [4], *Abdelwaheb Akremi* [2] and *Mathieu Abel* [1]

[1] Aix Marseille Université, CNRS, IM2NP, UMR 7334, Campus de St Jérôme, 13397, Marseille, France

[2] Université de Carthage, Faculté des Sciences de Bizerte, Laboratoire de Physique des Matériaux: Structure et Propriétés, LR01 ES15, Unité de Service Commun Spectromètre de Surfaces, 7021 Bizerte, Tunisie

[3] Centre d'élaboration de matériaux et d'études structurales (CEMES), CNRS, Université de Toulouse, Toulouse, France

[4] Aix Marseille Université, CNRS, CINAM, UMR 7325, Campus de Luminy, 13288, Marseille, France

**S1 Experimental methods**

**S2 Additional STM images**

**S3 LEED characterization**

**S4 XPS investigation**

**S5 Computational details**

**S6 Ab-initio results for $Mn_3C_6O_6$**

**S7 Ab-initio results for $Cu_3C_6O_6$**

## S1 Experimental methods

Experiments are performed in an ultra-high vacuum (UHV) system from Omicron instruments. The sample can be transferred in UHV between a preparation and an analysis chamber equipped with electron, X-ray sources hemispherical analyzer and STM microscopy. The base pressure of the analysis chamber is typically around $1 \times 10^{-10}$ mbar. The sample is a single crystal cut from Ag (111) wafer. The Ag(111) sample was prepared via multiple cycles of Ar$^+$ sputtering (0.8 keV) and annealing (up to 800 K) in a preparation chamber to recover the crystalline surface and therefore the surface order. The THQ was deposited onto the Ag(111) sample. A typical THQ-Ag island formed after deposition is shown in Figure S1. The materials that were used were copper and manganese as metal source evaporated from an EFM evaporator, combined with tetrahydroxyquinone (THQ). The ligand precursor THQ is deposited from an evaporator crucible and is let to react on Ag (111) with the metal atoms. Surface cleanliness and long-range order were confirmed in all experiments by X-ray photoelectron spectroscopy (XPS).

X-ray photoelectron spectroscopy (XPS) measurements were carried out using a monochromatic Al Kα X-ray source (photon energy 1486.6 eV, anode operating at 15 kV, Omicron) as incident radiation and a hemispherical electron spectrometer (Omicron EA 125). All scanning tunneling microscopy (STM) experiments were performed in a commercial (Omicron VT-STM) chamber operated in ultrahigh vacuum. STM images were collected with a specially prepared tip and WSXM software [1] was used to process all STM images. Low energy electron diffraction LEED (Omicron SPECTA-LEED) measurements were performed at room temperature.

**S2 Additional STM images**

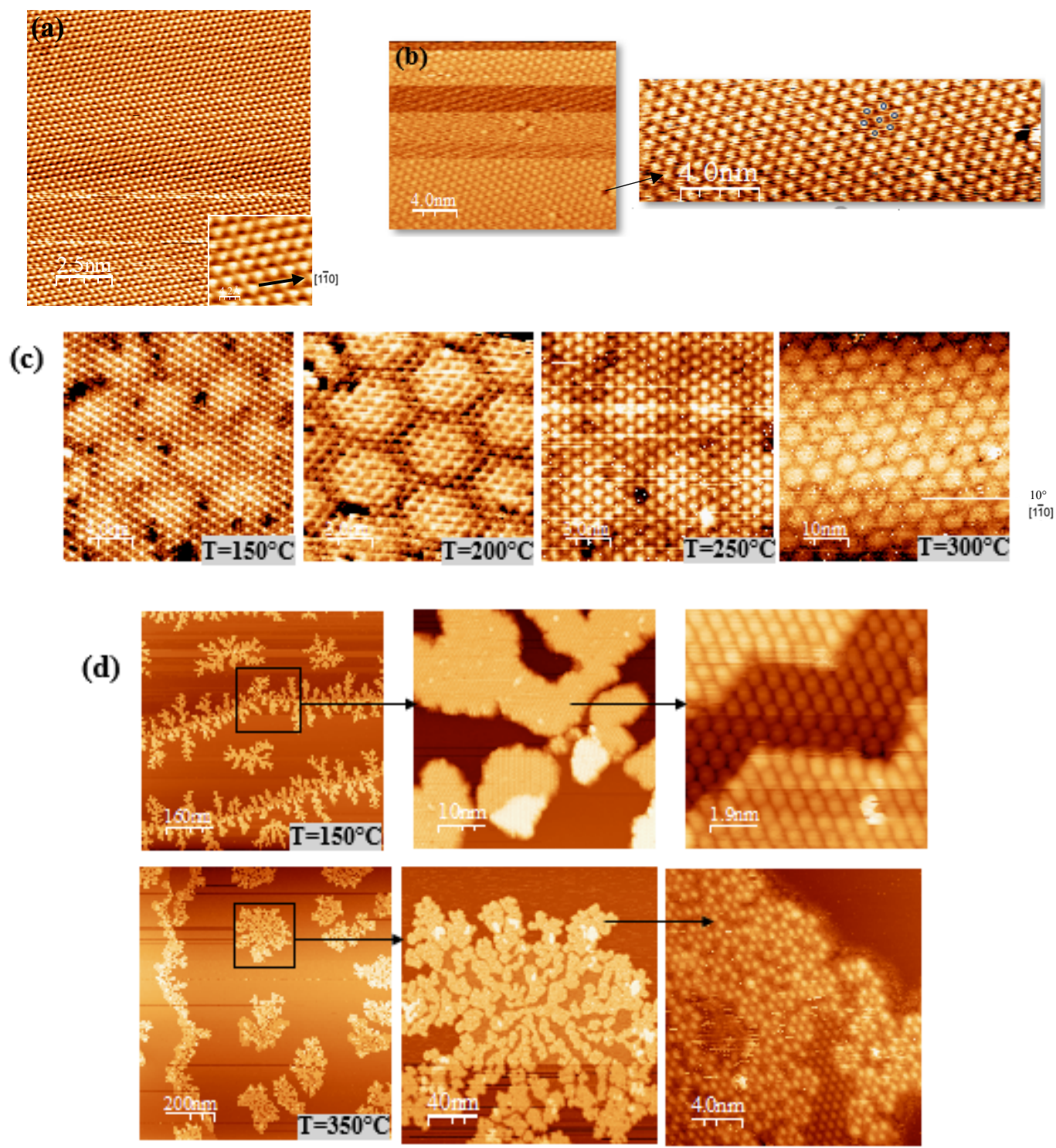

**Figure S1.** STM image of the Ag(111) surface recorded before (a) and (b) after 15 min of THQ deposition time. STM images acquired after co-depositing (c) (THQ-Cu) (d) (THQ-Mn) molecules on a Ag(111) substrate held at different temperatures.

Figure S1(b) displays THQ-Ag islands that formed after depositing THQ on Ag(111) sample. The image shows the regular hexagonal structure of THQ, highlighted by black points on Figure S1(b), the inset is 4 nm large. Figure S1(c) is a STM image after THQ+Cu co-deposition at different temperatures in which we observe Moiré patterns. Multi-domain islands are formed when co-depositing THQ and Mn atoms on a Ag(111) sample.

**S3 LEED characterization**

LEED controls were carried out after each STM analysis. Figure S2(a) shows the LEED patterns recorded after cleaning procedure of the Ag (111) sample and (b) after 15 min of THQ deposition. At first this LEED pattern is in line with a THQ plane in epitaxy on the Ag(111) substrate. Indeed, a slight rotation of THQ spots indicates small rotations of the THQ layer with respect to the substrate.

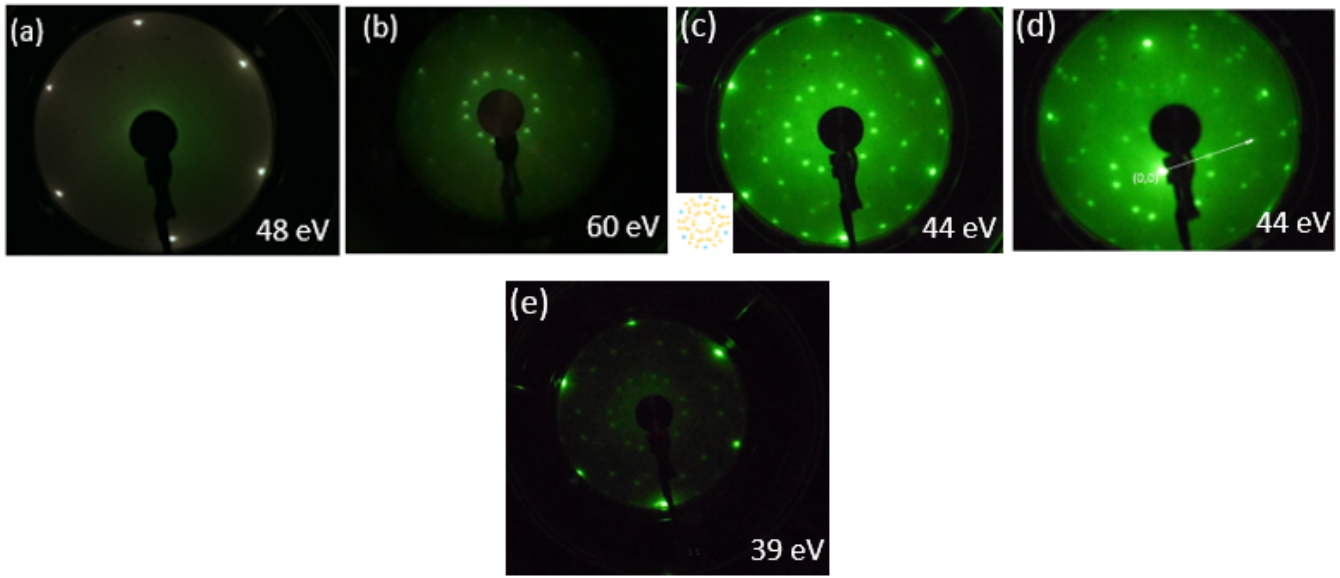

**Figure S2.** Low energy diffraction patterns, obtained for (a) the Ag(111) surface (b) after 1 ML THQ shown Ag(111)- (√7×√7)R20°-THQ, after co-deposit Cu and THQ (c,d) and Mn and THQ (e) on Ag(111).

Figure S2(c) shows the LEED pattern of the Ag(111) surface after the co-deposition of Cu and THQ. In comparison with Figure S2(d) the sample holder has been rotated a few degrees in order to show the extra spots around the (00) Ag spot. Note that the ratio of the spot distances measured on LEED is about 1.05 which corresponds to the ratio between THQ and THQ+Cu interatomic distances (8.1/7.6 =1.06). That is detailed in Fig. S3.

LEED pattern observed after co-deposition of Mn and THQ on Ag(111) ($E_P$= 39 eV) is shown in Figure S2(e). In the case of manganese, we have observed two domains in the STM image. On the LEED image we do not see two spots as in the case of copper and we conclude that the angle is small between Mn+THQ and THQ/Ag(111). The matrix form $M = \begin{pmatrix} 2 & 1 \\ -1 & 3 \end{pmatrix}$ is included for the $(\sqrt{7} \times \sqrt{7}) - R20°$ periodicity in Fig S3(b).

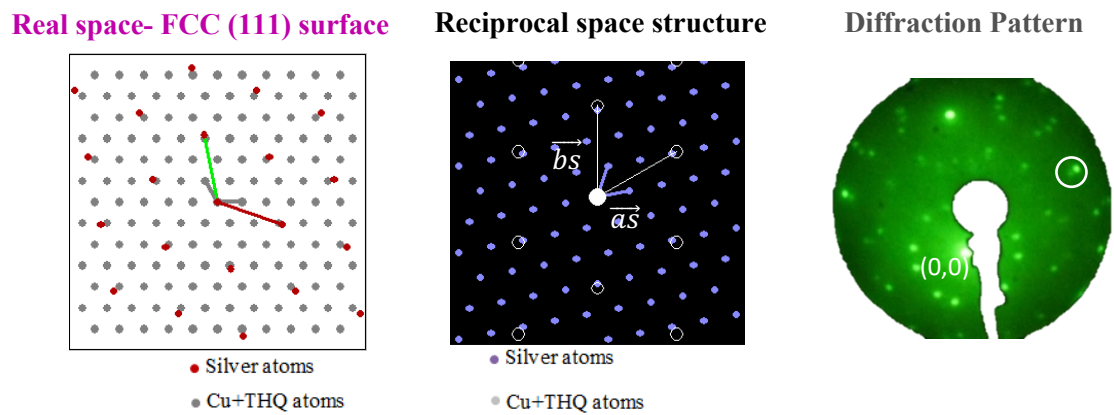

**Real space- FCC (111) surface**　**Reciprocal space structure**　**Diffraction Pattern**

• Silver atoms
• Cu+THQ atoms

**Figure S3.** Schematic model of the direct (a) and (b) reciprocal lattice for $(\sqrt{7} \times \sqrt{7})$ R20° reconstruction (c).

## S4 XPS investigation

X-ray photoemission spectroscopy allows to investigate the core levels of the atoms present at the surface, giving information on the chemical composition of the first surface layers. Furthermore, the analysis of core-level shifts provides precious information on the chemical environment, while the quantitative analysis of the peak intensities gives a reliable indication on the adsorbate coverage.

**(a)**

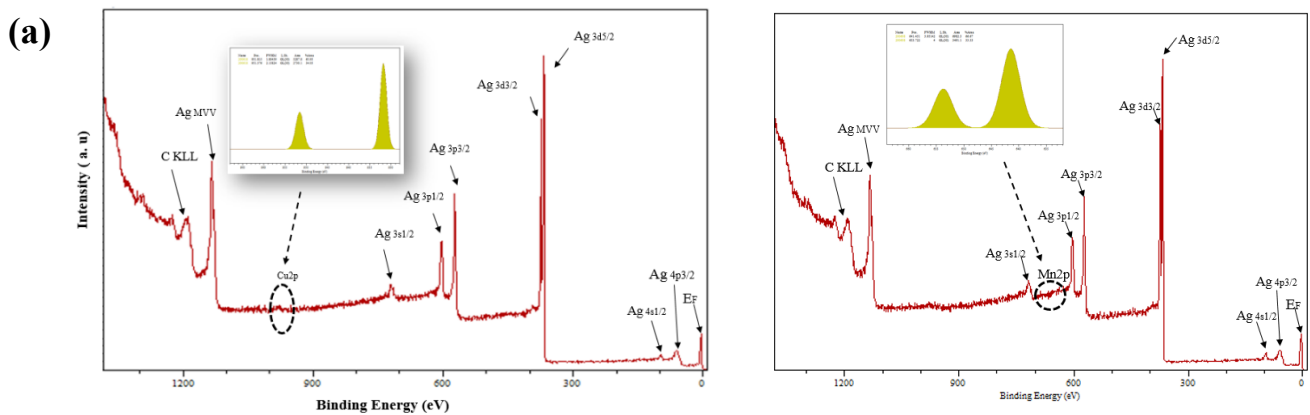

**(b)**

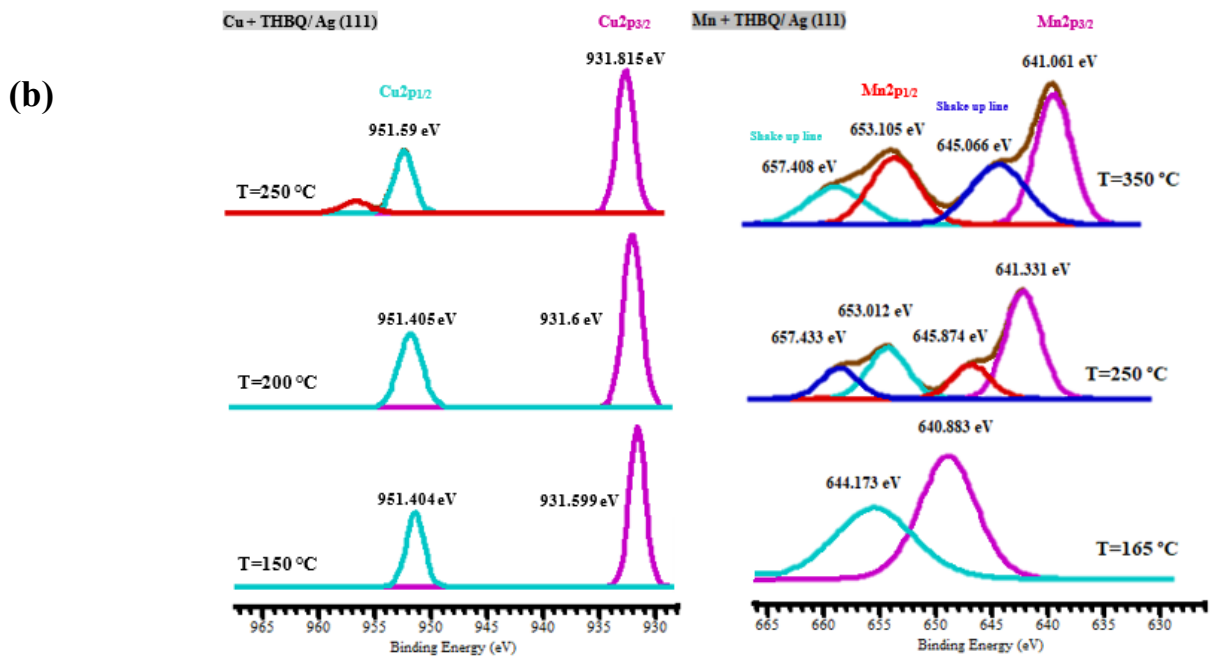

**(c)**

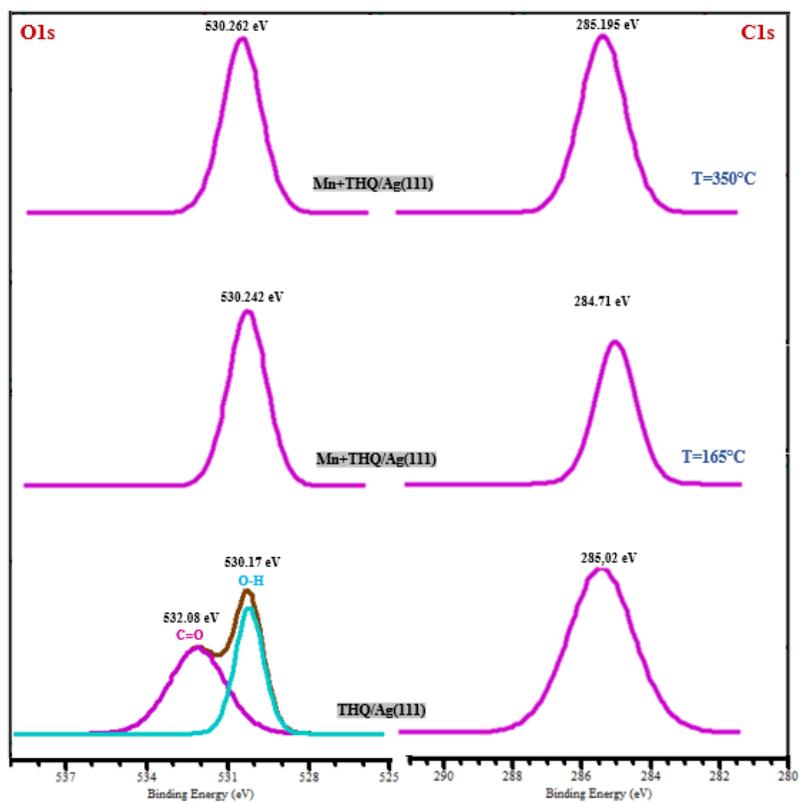

**Figure S4. (a)** XPS survey scan of the sample after co-depositing. **(b)** XPS Mn 2p and Cu 2p core level peaks. **(c)** XPS O 1s and C 1s core level peak evolutions before and after co-deposit at different temperatures.

Figure S5 is a ball model of a Cu and THQ monolayer on the Ag(111) plane. The observed Moiré pattern corresponds either to a 5 x 3 or a 4 x 4 superstructure of the Cu₃C₆O₆ unit. Correspondingly, the Moiré phase exhibits three possible lattice orientations that are about 2.5° (red line), 10.3° (blue line) and 18.1° (white line) relative to the [1-10] directions of Ag (111) (Table S1).

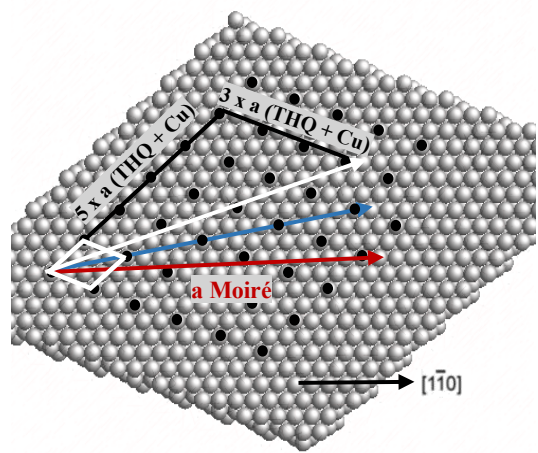

**Figure S5.** Ball model of a dense plane of Cu and THQ on Ag (grey balls are the silver atoms). The unit cell of the Moiré is highlighted in white.

**Table S1:** The Moiré phase periodicities and orientations with respect to the [1-10] direction.

| Θ (°) | a moiré (Å) | a THQ + Cu (Å) |
|---|---|---|
| 2.5 | 56.410 | 8.058 |
| 10.3 | 55.813 | 8.056 |
| 18.1 | 56.256 | 8.037 |

## S5 Computational details

The first-principle calculations were based on spin-polarized density functional theory (DFT) using the generalized gradient approximation (GGA) in the form proposed by Perdew, Burke, and Ernzerhof (PBE) as implemented in the Vienna ab initio simulation package (VASP) code. [2] For magnetic systems with an incompletely filled shell of 3d electrons, we use the spin polarized generalized gradient approximation with Hubbard term $U$ (SGGA+U). [3–5] We study isolated monolayers of $M_3C_6O_6$ (M=Cu or Mn) as well as monolayers adsorbed on the Ag(111) surface. The calculations determine the atomic positions, the magnetic couplings and the electronic structure. We used a correlation energy ($U$) of 5 eV and an exchange energy ($J$) of 0.90 eV for Mn and Cu d orbitals; these values have been tested and used in previous theoretical works for metal-organic polymers. [6–9] We also tested $U = 4$ eV for these materials and obtained results quite similar to those for $U = 5$ eV. The projected augmented wave (PAW) method [10,11] with a plane-wave basis set was used. We applied periodic boundary conditions and a vacuum space of 20 Å along the z direction both for the isolated or the adsorbed monolayer. Monkhorst and Pack [12] developed a common recipe for the k-point grid, and we use here a grid of (6×6×1) for the strictly two-dimensional (2D) materials. The smearing was set to 0.01 eV and the energy cutoff to 480 eV. The convergence criteria for ionic steps were set to $10^{-6}$ eV/Å and that one for electronic steps to $10^{-7}$ eV, which is sufficient for the structures studied here. The geometry optimization was performed with the variable lattice parameter and the full relaxation of the coordinates. To investigate the magnetic anisotropy energy (MAE), we apply the SGGA+U method in combination with spin orbit coupling (SOC). For the transition metals, three different magnetization directions are considered under SOC calculations, i.e., x, y, and z direction.

For the adsorbed monolayers on Ag(111) we took 6 Ag-layers into account and relaxed the uppermost three layers together with the $M_3C_6O_6$ layer. For the remaining atomic positions, we used the experimental lattice constant of Ag or an extended lattice to simulate the Moiré pattern as explained in the main text. For metal-organic monolayer with substrate we use here a grid of (4×4×1). The structures with substrate were relaxed until the residual forces were smaller than 0.02 eV/Å and we used a first order Methfessel-Paxton setting with a smearing (SIGMA) of 0.2 eV. The method of Methfessel-Paxton also results in a very accurate description of the total energy. Moreover, the Van-der-Waals dispersion correction term of DFT-D3 method proposed by Becke-Jonson was also adopted. [13] The positions of all atoms in the structures and the first three Ag layers were relaxed vertically and laterally.

## S6 Ab-initio results for Mn$_3$C$_6$O$_6$

The crystal, electronic and magnetic structures of free-standing monolayers of Mn$_3$C$_6$O$_6$ are shown in Figures S6 and S7. We investigated several magnetic structures (one ferromagnetic and two antiferromagnetic structures, Figure S1) to estimate the magnetic exchange couplings to first and second neighbors in the Kagomé lattice of magnetic centers.

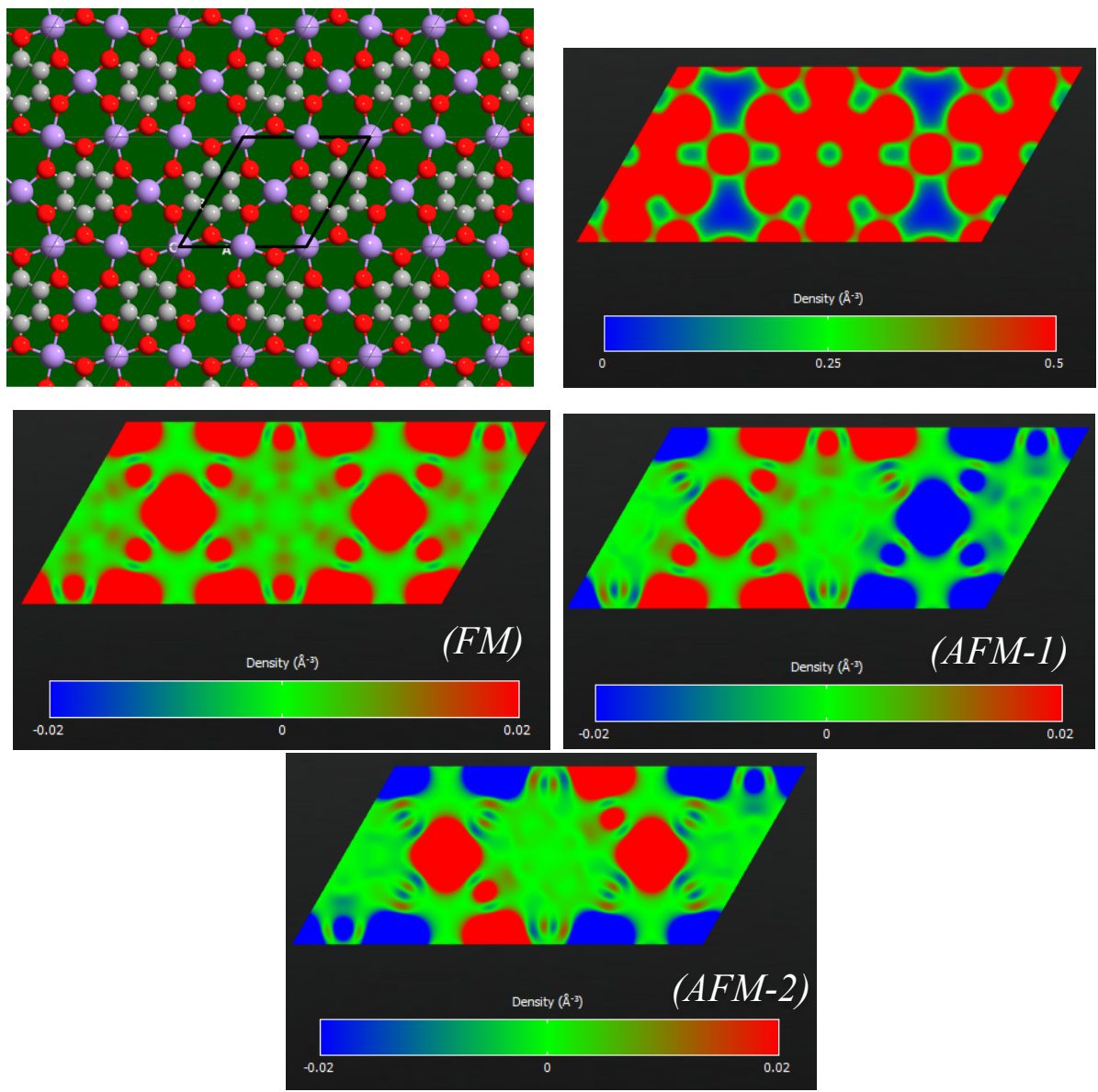

**Figure S6.** Geometrical structure, charge density and spin density of a free-standing monolayer of Mn$_3$C$_6$O$_6$ (structure A with 4 equivalents nearest neighbor oxygens for each Mn, see main text) (C: gray; O: red; Mn: purple). The results were obtained with the SGGA+U functional.

**Table 1.** Energy difference $\Delta_{E_{HS}-E_{IS}} = E(HS) - E(IS)$ between HS and IS states per unit cell in eV, distance between the Mn atoms and the O atoms ($d_{Mn-O}$, in Å), lattice constant ($a$, in Å), exchange energies ($E_{ex-1/2} = E_{AFM\,(1/2\,)} - E_{FM}$ per Mn atom, in meV), exchange coupling constants $J_1$ and $J_2$ as explained in the text in meV , total magnetic moments ($M$ per Mn atom, in $\mu_B$), local magnetic moments of the d orbital at the Mn atoms ($M_d$ per Mn atom, in $\mu_B$), local magnetic moments at the Mn atoms ($M_m$ per Mn atom, in $\mu_B$), energy band gaps (spin up ($E_a$) and spin down ($E_b$), in eV), general energy gaps ($E_g$, in eV), for 2D $Mn_3C_6O_6$.

| Mn₃C₆O₆ | | | |
|---|---|---|---|
| $U$ | **Without** | **4 eV** | **5 eV** |
| $\Delta_{E_{HS}-E_{IS}}$ | -1.96 | -4.24 | -4.38 |
| $d_{Mn-O}$ | 2.10 | 2.10 | 2.10 |
| $a$ (Å) | 7.941 | 7.968 | 7.975 |
| $E_{ex-1}$ | -26.07 | -8.56 | -6.08 |
| $E_{ex-2}$ | -46.35 | -15.38 | -10.87 |
| $J_1$ | 2.43 | 0.82 | 0.57 |
| $J_2$ | 0.35 | 0.10 | 0.08 |
| $M$ | 5 | 5 | 5 |
| $M_d$ | 4.363 | 4.531 | 4.576 |
| $M_m$ | 4.445 | 4.608 | 4.651 |
| $E_a$ | 0.64 | 1.24 | 1.22 |
| $E_b$ | 1.08 | 1.42 | 1.50 |
| $E_g$ | 0 | 1.20 | 1.22 |

**Table S2:** Magnetic anisotropy energy (MAE, per Mn atom in meV), total magnetic moments $M_i$ local magnetic moments of the d orbitals at the Mn site $M_d(i)$ and orbital moments ($M_L(i)$) per Mn atom in $\mu_B$) for Mn$_3$C$_6$O$_6$ (without substrate). The data result from SGGA-PBE and SGGA+U calculations including SO coupling with the magnetic moments directed towards the $i$={$x,y,z$} direction.

| Mn$_3$C$_6$O$_6$ | | | |
|---|---|---|---|
| U | Without | 4 eV | 5 eV |
| MAE ($E_x$-$E_z$) | -0.35 | -0.39 | -0.33 |
| MAE ($E_x$-$E_y$) | 0 | 0 | 0 |
| $M_x$ | 5 | 5 | 5 |
| $M_y$ | 5 | 5 | 5 |
| $M_z$ | 5 | 5 | 5 |
| $M_d(x)$ | 4.362 | 4.528 | 4.573 |
| $M_d(y)$ | 4.362 | 4.528 | 4.573 |
| $M_d(z)$ | 4.362 | 4.529 | 4.574 |
| $M_L(x)$ | 0.002 | 0.002 | 0.002 |
| $M_L(y)$ | 0.002 | 0.002 | 0.002 |
| $M_L(z)$ | 0.003 | 0.002 | 0.002 |

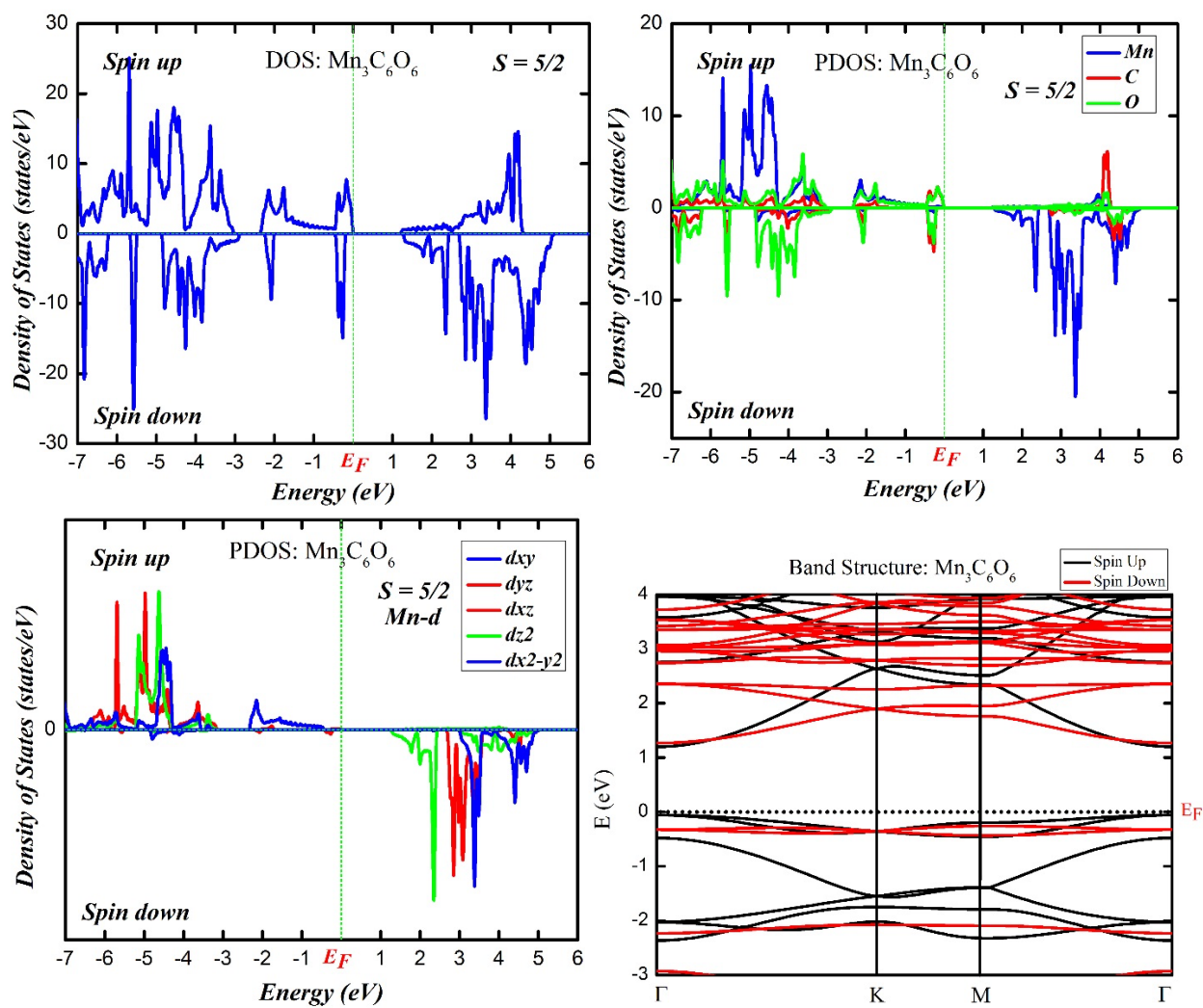

**Figure S7.** Total, projected DOS and band structure of the Mn, C, O atoms and d orbitals on the Mn atoms in an isolated monolayer of $Mn_3C_6O_6$ without charge transfer with the SGGA+U-PBE method.

The crystal structure, as well as the charge and magnetization densities of $Mn_3C_6O_6$ including substrate are shown in Figure S8. For the corresponding densities of states, please see Figure S9. We also calculated the adsorption characteristics, i.e. the vertical distances between the atoms of the monolayer and the uppermost Ag(111) plane (Table S3). The adsorption energies are significantly increased if we take into account the van-der-Waals interaction on the level of the PBE-D3 functional. [13] It is remarkable that the adsorption energy increases by a factor of two roughly and the corresponding distances between monolayer and noble atom surface reduce by taking into account the van-der Waals interaction. Also, the Mn-atoms are much closer to the Ag-surface than the C and O atoms. Finally, the information contained in the Bader charge analysis (Table S4) is crucial since it defines the charge transfer between substrate and monolayer.

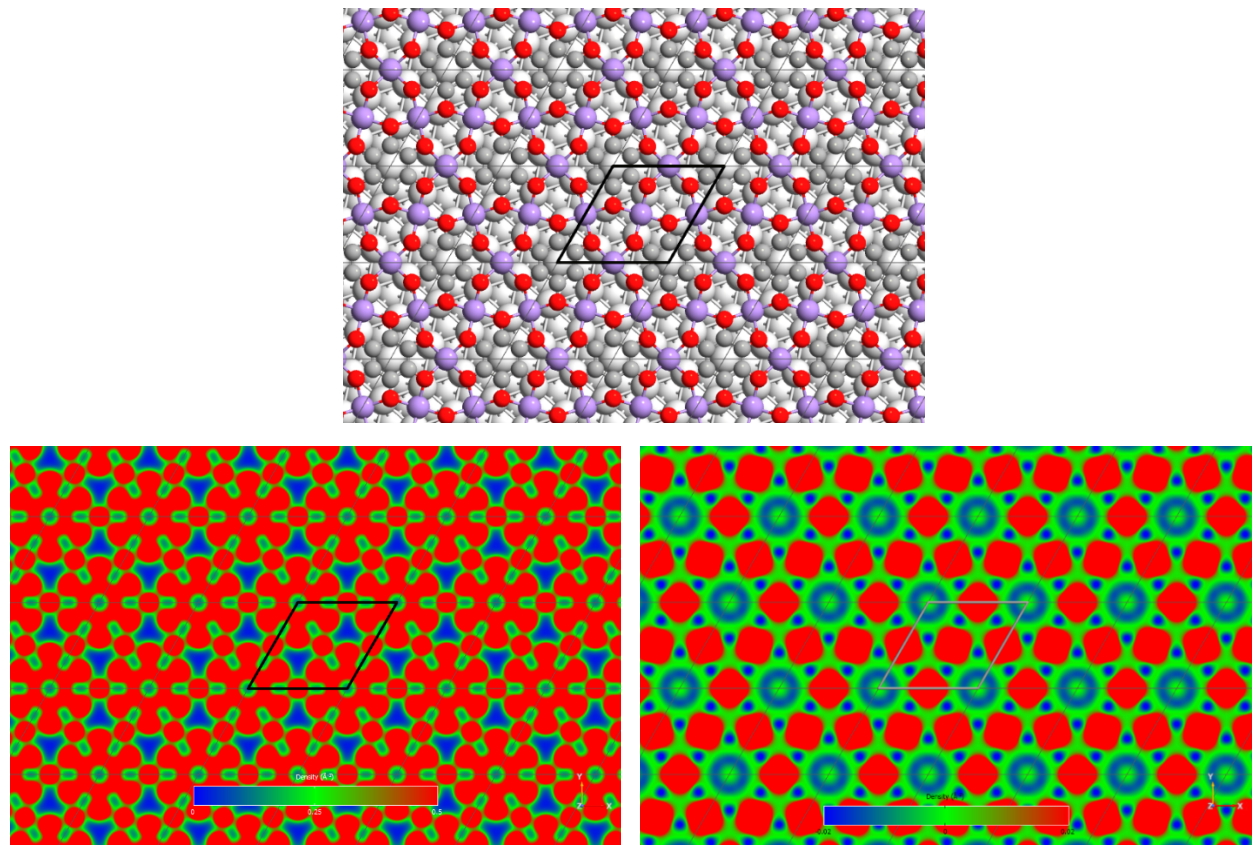

**Figure S8.** Geometrical structure, charge and spin densities of $Mn_3C_6O_6$ on the Ag(111) surface. The spin density varies between -0.02 and 0.02 Å$^{-3}$ for the cut plane.

**Table S3:** Distance between the Mn atoms and the O atoms ($d_{Mn-O}$, in Å), vertical distances between the Mn, C, O atoms and the Ag(111) substrate ($d_{Mn-Ag(111)}$, $d_{C-Ag(111)}$ and $d_{O-Ag(111)}$, in Å), total magnetic moments ($M$ per cell, in $\mu_B$), local magnetic moments of the d orbital at the Mn atoms ($M_d$ per Mn atom, in $\mu_B$), local magnetic moments at the Mn atoms ($M_m$ per Mn atom, in $\mu_B$), adsorption energies ($E_{ads}$ per cell in eV), for $Mn_3C_6O_6$ on the Ag(111) surface.

| | **Without U** | **U = 5 eV** | **PBE-D3** | **PBE+U-D3** |
|---|---|---|---|---|
| $d_{Mn-O}$ | 2.07 | 2.13 | 2.07 | 2.13 |
| $d_{Mn-Ag(111)}$ | 2.61 | 2.61 | 2.46 | 2.49 |
| $d_{C-Ag(111)}$ | 3.37 | 3.59 | 3.20 | 3.44 |
| $d_{O-Ag(111)}$ | 3.16 | 3.34 | 3.01 | 3.20 |
| $M$ | 13.27 | 13.67 | 13.23 | 13.67 |
| $M_d$ | 4.11 | 4.54 | 4.08 | 4.53 |
| $M_m$ | 4.17 | 4.61 | 4.15 | 4.60 |
| $E_{ads}$ | -1.90 | -1.84 | -3.87 | -3.71 |

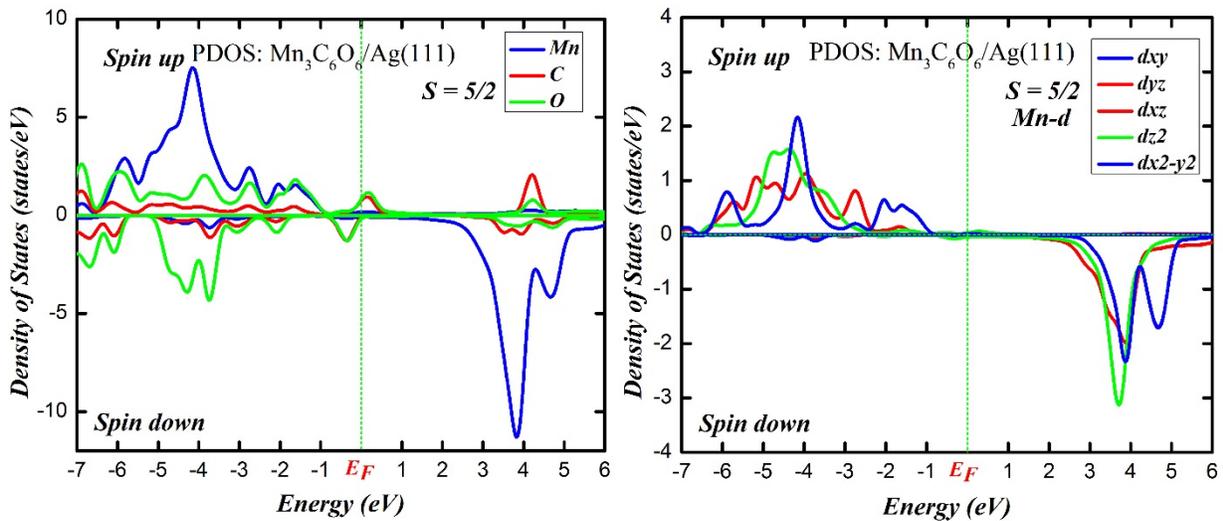

**Figure S9.** Projected DOS of the Mn, C, O atoms and of the d orbitals on the Mn site of $Mn_3C_6O_6$ deposited on the Ag(111) surface in the FM spin configuration.

**Table S4:** Bader charge distribution for isolated and adsorbed monolayers Mn$_3$C$_6$O$_6$.

| | *Mn$_3$C$_6$O$_6$* | *Mn$_3$C$_6$O$_6$/Ag(111)* |
|---|---|---|
| Mn | 5.5673 | 5.6756 |
| Mn | 5.5630 | 5.6766 |
| Mn | 5.5607 | 5.6768 |
| C | 3.4743 | 3.4116 |
| C | 3.5000 | 3.3772 |
| C | 3.3262 | 3.2568 |
| C | 3.6451 | 3.3698 |
| C | 3.4989 | 3.4300 |
| C | 3.5031 | 3.5397 |
| O | 7.2169 | 7.1667 |
| O | 7.2195 | 7.1705 |
| O | 7.2188 | 7.1681 |
| O | 7.2240 | 7.1678 |
| O | 7.2418 | 7.1608 |
| O | 7.2404 | 7.1653 |
| sum | 81 | 80.4133 |
| **Charge-transfer** | **-0.59 e** | |

## S7 Ab-initio results for Cu₃C₆O₆

The results for isolated monolayers are summarized in Figures S10 (crystal structure, charge and spin densities) and S11 (DOS). Figures S12 (atomic positions and charge density) and S13 (DOS) concern the adsorbed monolayer, as well as Tables S5 (adsorption energies) and S6 (Bader charge analysis).

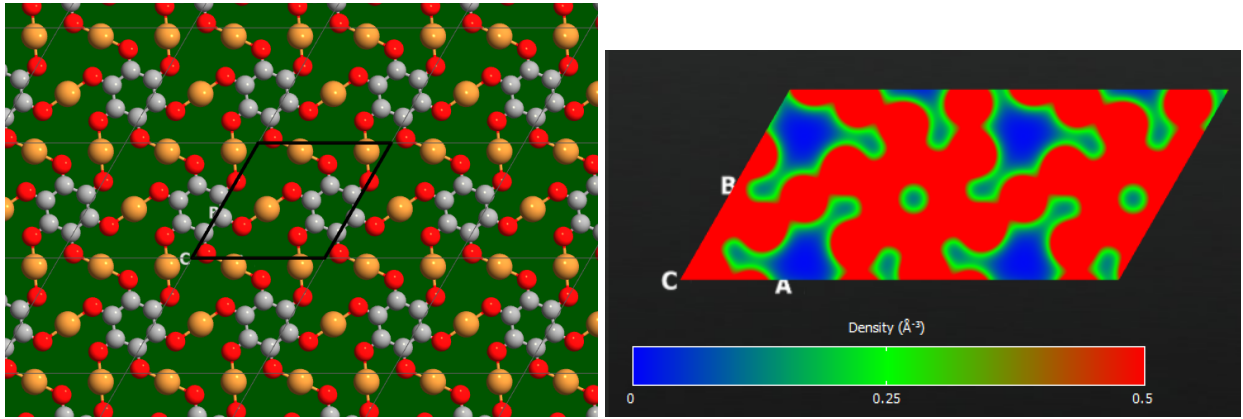

**Figure S10.** Geometrical structure and charge density of the Cu₃C₆O₆ monolayer.

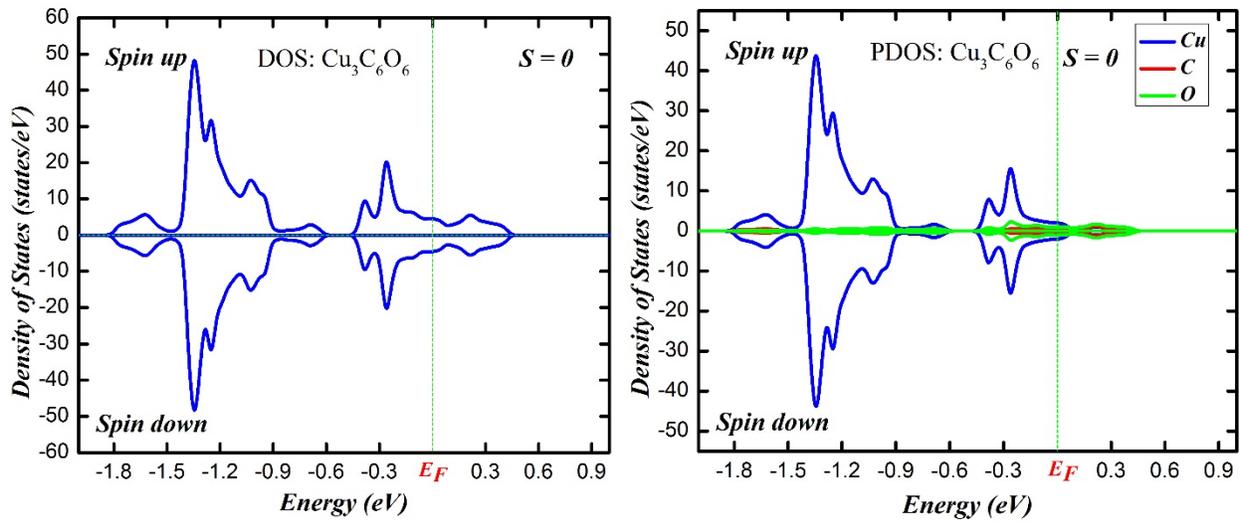

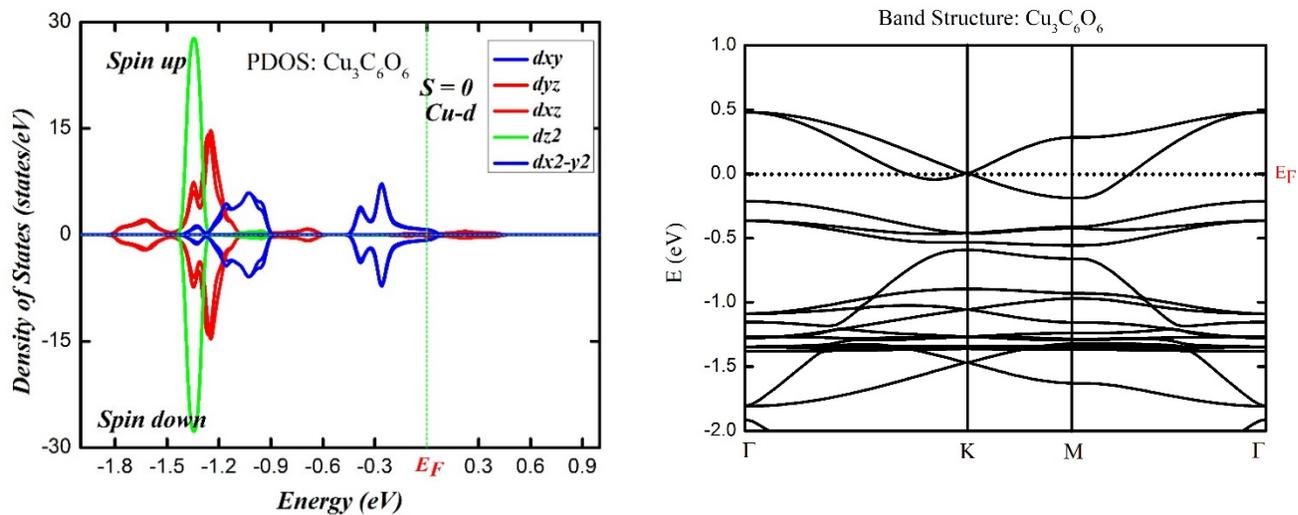

**Figure S11.** Total, projected DOS onto the Cu, C, O atoms and d orbitals of Cu and Band Structure in the strictly 2D Cu₃C₆O₆.

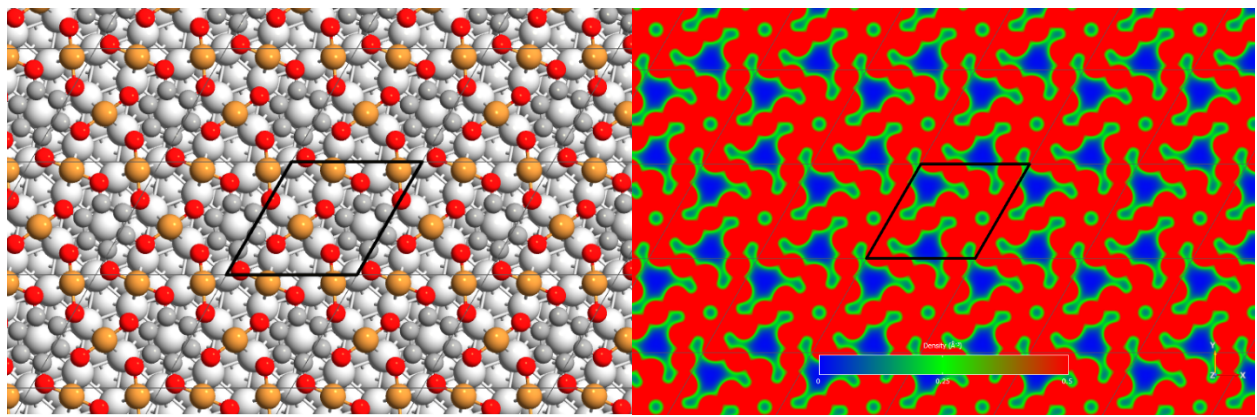

**Figure S12.** Geometrical structure and charge density of $Cu_3C_6O_6$ on the Ag(111) surface.

**Table S5:** Distance between the Cu atoms and the O atoms ($d_{Cu\text{-}O}$, in Å), vertical distance between the Cu, C, O atoms and the Ag(111) substrate ($d_{Cu\text{-}Ag(111)}$, $d_{C\text{-}Ag(111)}$ and $d_{O\text{-}Ag(111)}$ , in Å), adsorption energies ($E_{ads}$ per cell in eV), for 2D $Cu_3C_6O_6$ on the Ag(111) Surface.

| | **GGA** | **SGGA** | **SGGA-D3** |
|---|---|---|---|
| $d_{Cu\text{-}O}$ | 1.89 | 1.89 | 1.89 |
| | 2.50 | 2.50 | 2.48 |
| $d_{Cu\text{-}Ag(111)}$ | 2.37 | 2.37 | 2.30 |
| $d_{C\text{-}Ag(111)}$ | 2.82 | 2.82 | 2.75 |
| $d_{O\text{-}Ag(111)}$ | 2.72 | 2.72 | 2.66 |
| $M$ | 0 | 0 | 0 |
| $E_{ads}$ | -1.28 | -1.28 | -3.32 |

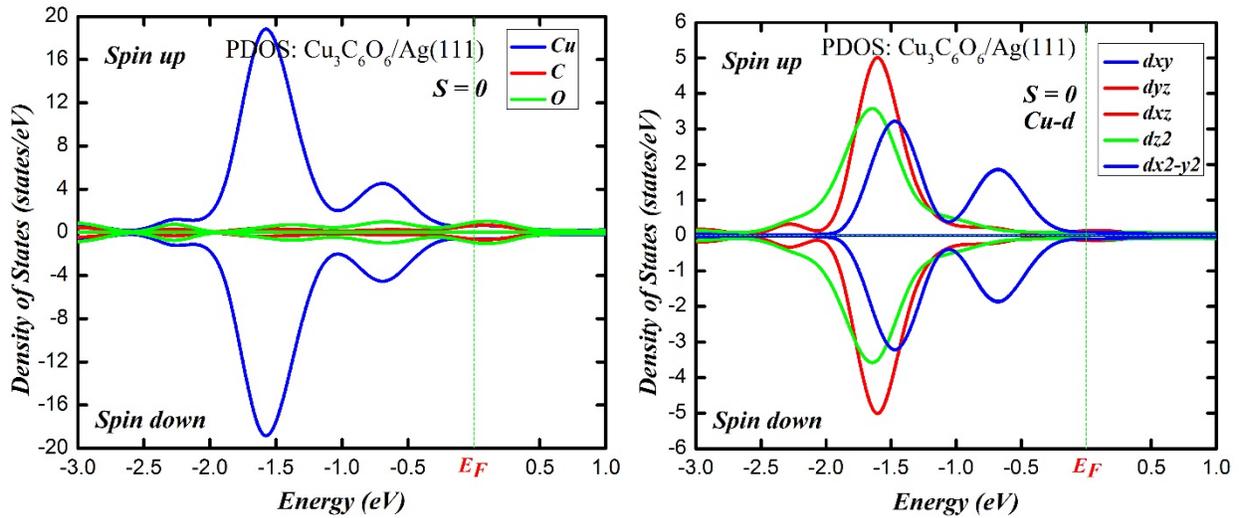

**Figure S13.** Projected DOS of the Cu, C, O atoms and of the d orbitals on the Cu atom in $Cu_3C_6O_6$ deposited on the Ag(111).

**Table S6:** Bader charge distribution for all atoms in isolated and adsorbed $Cu_3C_6O_6$.

| | $Cu_3C_6O_6$ | $Cu_3C_6O_6/Ag(111)$ |
|---|---|---|
| Cu | 10.2203 | 10.2656 |
| Cu | 10.2186 | 10.2658 |
| Cu | 10.2193 | 10.2634 |
| C | 3.2698 | 3.3750 |
| C | 3.3038 | 3.3125 |
| C | 3.3424 | 3.3338 |
| C | 3.3245 | 3.3331 |
| C | 3.3105 | 3.3720 |
| C | 3.3002 | 3.4115 |
| O | 7.0844 | 7.0798 |
| O | 7.0832 | 7.0897 |
| O | 7.0767 | 7.0894 |
| O | 7.0723 | 7.0759 |
| O | 7.0929 | 7.0683 |
| O | 7.0810 | 7.0935 |
| sum | 93 | 93.43 |
| **Charge-transfer** | 0.43 e | |